\documentclass[preprint,bibnotes,prb,amsmath,endfloats*,preprintnumbers,amssymb,amsfonts]{revtex4}
\usepackage{mathpazo}
\usepackage{graphicx}
\begin{document}
\title{Carrier-mediated magnetoelectricity in complex oxide heterostructures}
  \author{James M.\ Rondinelli}
  \author{Massimiliano Stengel}
  \author{Nicola A.\ Spaldin}
     \email[Address correspondence to: ]{nicola@mrl.ucsb.edu}
     \thanks{\textit{Competing Interests.} The authors declare that they have             no competing financial interests.}
  \affiliation{Materials Department, University of California, Santa Barbara, CA, 93106-5050, USA}
\date{\today}

\begin{abstract}
The search for a general means to control the coupling between
electricity and magnetism has intrigued scientists since {\O}rsted's
discovery of electromagnetism in the early 19$^{\rm th}$ century.
While tremendous success has been achieved to date in creating both
single phase and composite magnetoelectric materials, the quintessential 
electric-field control of magnetism remains elusive.
In this work, we demonstrate a  linear magnetoelectric effect which 
arises from a novel carrier-mediated mechanism, and is a universal feature of the 
interface between a dielectric and a spin-polarized metal.
Using first-principles density functional calculations, we illustrate this effect at
the SrRuO$_3$/SrTiO$_3$ interface and describe its origin.
To formally quantify the magnetic response of such an interface to an applied electric
field, we introduce and define the concept of spin capacitance.
In addition to its magnetoelectric and spin capacitive behavior, the interface 
displays a spatial coexistence of magnetism and dielectric polarization suggesting 
a route to a new type of interfacial multiferroic. 
\end{abstract}
\maketitle

The linear magnetoelectric effect is the first-order magnetic response
of a system to an applied electric field, or equivalently the electrical
polarization induced by an applied magnetic field \cite{ODell:1970,Fiebig:2005}:
\begin{eqnarray}
P_i & = & \alpha_{ij} H_j \\
M_i & = & \alpha_{ji} E_j \quad,
\end{eqnarray}
where $\alpha$ is the magnetoelectric tensor, which is non-zero in the
absence of time-reversal and space-inversion symmetries.
Although the effect has long been recognized both theoretically
\cite{Dzyaloshinskii:1960} and experimentally \cite{Astrov:1960}
it has traditionally been of only academic interest because
of its inherent weakness in intrinsic single phase materials. 
The last few years, however, have seen a tremendous revival of activity 
in magnetoelectrics \cite{Fiebig:2005}, motivated in large part by the 
entirely new device paradigms that would be enabled by electric-field-control 
of magnetism \cite{Binek/Doudin:2005,Borisov_et_al:2005}. 
Progress has been fueled by the growth of novel single-phase 
\cite{Wang_et_al:2003,Kimura_et_al_Nature:2003}
and composite \cite{Srinivasan_et_al:2001,Zheng_et_al:2004} multiferroic 
materials, which show simultaneous magnetic and ferroelectric ordering 
and have the potential for significantly enhanced magnetoelectric responses. 

Currently two mechanisms for magnetoelectric coupling are well established.
In traditional single-phase magnetoelectrics, of which Cr$_2$O$_3$ is the
prototype, an {\it intrinsic} magnetoelectric coupling occurs \cite{Dzyaloshinskii:1960}. An electric 
field both shifts the positions of the magnetic cations relative to the anions --
changing both the dipolar and exchange contributions to the magnetic interactions
-- and modifies the electronic wavefunctions, which further
changes the magnetic coupling. This behavior, and its inverse when a
magnetic field is applied, are enhanced in materials
with strong spin-orbit coupling (and hence strong coupling of the magnetic
order to the lattice) and with large piezo- or electro-strictive response.
In composites of magnetostrictive or piezomagnetic materials combined with
electrostrictive or piezoelectrics, an {\it extrinsic} magnetoelectric coupling 
is mediated by strain at the interface \cite{Srinivasan_et_al:2001,Dong_et_al:2003}. 
Here an electric field causes strain in the electrical component which is 
mechanically transferred to the magnetic component where it changes the magnetization 
(and vice versa)
\footnote{
Although not strictly a linear magnetoelectric effect, a related behavior 
in single-phase multiferroics, in which electric field re-orientation of 
ferroelectric polarization reorients the magnetic easy axis or plane 
through the elastic coupling \cite{Zhao_et_al:2006} and vice versa 
\cite{Kimura_et_al_Nature:2003}, has recently received considerable 
attention.}.
This class of structures appears very promising for technological applicability,
in part because the underlying working principles are essentially based on the 
bulk properties of the constituents.
Very accurate modeling has thus been achieved at the level of
semiclassical theories~\cite{Srinivasan_et_al:2001}, which has allowed
for efficient optimization of geometry and material properties.

Recently, it has been pointed out that thin film growth of heterostructures can 
be exploited to create structures which break inversion symmetry, even when the
parent bulk materials are centrosymmetric \cite{Yamada_et_al:2004}. 
In fact at the interface between any two unlike materials the space inversion 
symmetry is intrinsically broken. 
Therefore, provided that one of the two components breaks time-reversal symmetry 
(which occurs in the case of a ferromagnet), we should expect a linear magnetoelectric 
coupling to occur, even in the absence of a piezo- or electro-strictive material 
(Table~\ref{linearME}). 
An intriguing recent observation of electric-field tunable magnetization
at the surface of a ferromagnetic electrode in an electro-chemical cell
\cite{Weisheit_et_al:2007} may indeed be a manifestation of such an interfacial
effect.  Ref. \onlinecite{Weisheit_et_al:2007} suggested an important role played
by accumulation/depletion of carriers at the ferromagnetic surface,
but a detailed microscopic mechanism could not be determined.
A simpler and more convenient architecture for developing a quantitative and
general model is a capacitor geometry, in which one of the two components of 
the bilayer is a metal and the other is an insulating dielectric, and a second 
electrode is deposited on top of the insulating film to provide a source of 
external bias; we study this geometry here.

\begin{table}
\begin{center}
\begin{tabular}{l|ccccccc}
\vspace*{-6pt}
                           & magnetic   && non-magnetic && magnetic  && magnetic  \\
                           & non-polar  && polar        && polar     && interface \\ \hline 
time-reversal symmetric?   &  no        &&     yes      &&    no     &&  no       \\
space-inversion symmetric? &  yes       &&     no       &&    no     &&  no       \\
linear ME effect allowed?  &  no        &&     no       &&    yes    &&  yes      \\
\end{tabular}
\caption{The linear magneto-electric (ME) effect is only allowed in the absence of
both time-reversal and space-inversion symmetries. For a single phase material
the co-existence of electrical polarization and magnetic ordering are required
for linear magneto-electricity to occur;
in the presence of an interface which breaks space-inversion symmetry the existence
of magnetic ordering is sufficient.}
\label{linearME}
\end{center}
\end{table}

The magnetoelectric response of such an interface is intimately related to the local 
details of chemistry, bonding, structure and electronics,
which often strongly depart from those of single-phase compounds;
to account for the subtle balance of these often competing factors requires 
a fully quantum-mechanical treatment.
Density functional theory has proven to be a very effective tool to deal with the
challenging problem of complex oxide interfaces;
however, the lack of a rigorous methodological framework to treat finite electric 
fields in metal-insulator heterostructures has until recently thwarted attempts to 
access dielectric and magnetoelectric responses of such systems.
Recent advances in finite electric field methods 
\cite{Stengel/Spaldin:2007,Stengel/Spaldin_Nature:2006}, 
which we extended in this work to include the magnetic degree of freedom, enables such
calculations for the first time.

Here we demonstrate computationally a novel linear magnetoelectric effect which is
{\it carrier-mediated} in origin. The effect occurs at the interface between a 
non-magnetic, non-polar dielectric and a metal with spin-polarized carriers at
the Fermi level. 
Using first-principles density functional theory, we find that 
\begin{itemize}
\item[(i)] an electric field induces a linear change in magnetization in the metal at the
interface, with the sign of the change determined by the orientation of the 
applied field,
\item[(ii)] the magnetic response is mediated by the accumulation of spin-polarized 
charge within the few atomic layers closest to the interface; this charge is stored
capacitively leading to a spintronic analogue to a traditional charge
capacitor, and 
\item[(iii)] as a result of the magnetoelectric response, magnetism and dielectric 
polarization coexist in the interfacial region, suggesting a route to a new type
of {\it interfacial multiferroic}.
\end{itemize}


We choose SrRuO$_3$/SrTiO$_3$ heterostructures as our representative
system, motivated by the favorable dielectric properties
of SrTiO$_3$ (STO) \cite{Velev_et_al:2005} and the widespread use of thin film 
ferromagnetic SrRuO$_3$ (SRO) as oxide electrodes \cite{Marrec_et_al:2002}; 
furthermore, the effects of interfacial carrier modulation
in thin films of both materials have been intensively investigated as a possible
route to the realization of novel field-effect devices~\cite{Ahn_et_al:2006,
Takahasi_et_al:2006}.
We note that our conclusions are not specific to this system but should 
apply to any dielectric/magnetic metal interface. 
Since such interfaces are similar to those already in place for magnetic tunnel 
junctions \cite{Gallagher/Parkin_2006}, rapid experimental verification of this effect should be
well within reach.

\begin{figure}
\includegraphics[width=0.95\textwidth]{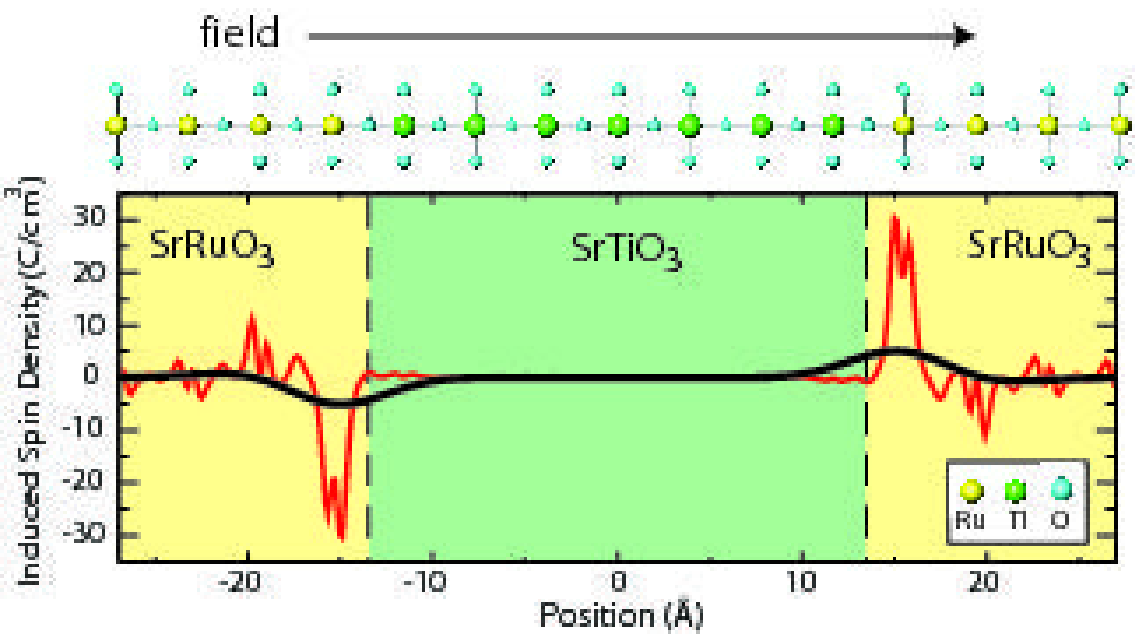}
\caption{(color online)~Calculated magnetization induced by an external voltage of 27.8 mV
in a nanocapacitor consisting of 7 layers of SrTiO$_3$ alternating with 7 layers
of ferromagnetic, metallic SrRuO$_3$. The light (red) and bold (black) lines
show respectively the planar averaged, and macroscopically and planar averaged induced 
magnetization profiles.} 
\label{spin_capacitor_fig}
\end{figure}

Our calculations are performed within the local spin-density approximation 
(LSDA) of density functional theory and the projector-augmented wave 
\cite{Bloechl:1994} (PAW) 
method~\footnote{ The PAW pseudopotentials were generated in 
the $4s^24p^65s^2$ configuration for Sr, $3s^23p^64s^23d^2$ for Ti, $2s^22p^4$ 
for O and $4s^24p^64d^75s^1$ for Ru.}, with a planewave basis cutoff energy 
of 40~Ry.
Our model heterostructure consists of seven unit cells of SrRuO$_3$ periodically 
alternating with seven unit cells of SrTiO$_3$ (Figure~\ref{spin_capacitor_fig}) along
the (100) direction.
The in-plane lattice parameter was set to the theoretical SrTiO$_3$ equilibrium lattice 
constant (3.85 \AA) to simulate epitaxial growth on a SrTiO$_3$ substrate.
The two-dimensional Brillouin zone of the resulting tetragonal cell was 
sampled with six special $k$-points, corresponding to a $6 \times 6$
Monkhorst-Pack mesh\cite{Monkhorst/Pack:1976}.
The ion positions and out-of-plane lattice constants were fully 
relaxed until the forces on the ions and the out-of-plane stress were 
less than 1 meV \AA$^{-1}$ using an extension of the Car and Parrinello method 
for  metallic systems \cite{Car/Parrinello:1985,Vandevondele/DeVita:1999,
Stengel/DeVita:2000}. 
To study dielectric and magnetoelectric responses we used a spin-polarized extension of
our recently developed method for applying finite electric fields to metal-insulator 
heterostructures \cite{Stengel/Spaldin:2007,Stengel/Spaldin_Nature:2006}.
With these parameters we obtain structural, dielectric and electronic properties for
bulk SrTiO$_3$ and SrRuO$_3$ that are consistent with earlier theoretical studies
\cite{Antons_et_al:2005,Allen_et_al:1996,Singh_SrRuO3:1996,Mazin/Singh:1997}. In particular, 
our calculated dielectric constant for cubic SrTiO$_3$ at this lattice constant is 
$\epsilon= 490$, in excellent agreement with Ref. \onlinecite{Antons_et_al:2005},
and our calculated magnetic moment for ferromagnetic SrRuO$_3$ epitaxially constrained 
to the LDA equilibrium lattice constant of SrTiO$_3$ (less than 0.2\% strain) is 
1.03 $\mu_B$ per Ru atom, consistent with 
Refs.\ \onlinecite{Allen_et_al:1996,Singh_SrRuO3:1996,Mazin/Singh:1997}.

To investigate the response of the heterostructure to an applied electric
field, we apply an external bias of $\Delta V=$27.8~mV across the capacitor 
plates, and allow the ions to relax to their equilibrium positions. 
The resulting
planar and macroscopically averaged change in magnetization is shown 
in Figure~\ref{spin_capacitor_fig}.
The overall induced magnetic moment is localized at the interfaces,
and amounts to $2.5 \times 10^{-3}$ $\mu_B$ per surface unit cell,
corresponding to a surface spin density of 0.27~$\mu$C/cm$^{2}$.
The accumulation of spin is exactly equal in magnitude and opposite in sign 
(within the numerical accuracy of our method) at the left and right electrode 
so that the overall induced magnetic moment of the heterostructure is zero, 
consistent with the symmetry of the system.
In the same Figure~\ref{spin_capacitor_fig} we show the planar average of the induced 
spin density without macroscopic averaging.
It is clear that the dominant contribution to the induced magnetization is accumulation
of spin on the interfacial RuO$_2$ layer; interestingly, this effect is partially 
compensated by a smaller {\it opposite} induced spin density in the adjacent RuO$_2$ 
layer, somewhat reminiscent of Kondo behavior \cite{Kondo}.
Conversely, the induced spin on the SrTiO$_3$ side of the interface, which is provided
by the exponentially vanishing tails of the metal-induced gap states, is small.

\begin{figure}
  \centering
  \includegraphics[width=0.950\textwidth]{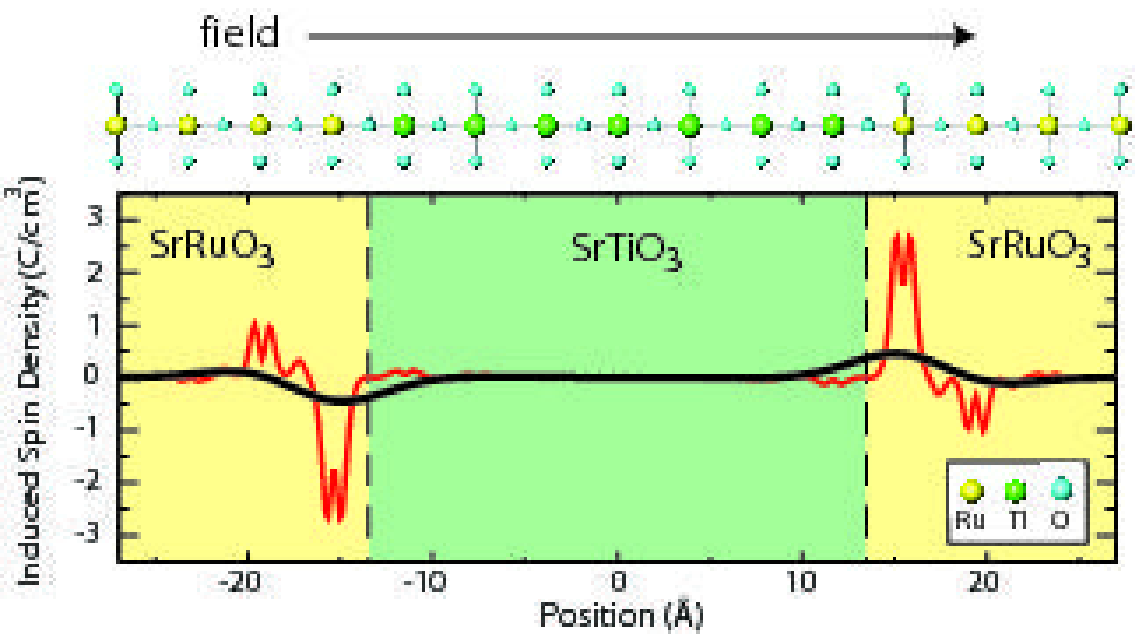}
  \caption{\label{hf_noavg}(color online)~Calculated high-frequency planar averaged 
  magnetization induced by an external voltage of 27.8~mV. 
  The light (red) and bold (black) lines 
  show respectively the planar averaged, and macroscopically and planar averaged 
  profiles. Note that the $y$-axis scale is exactly one order of magnitude 
  smaller than in the static case (Fig. 1). The strong change 
  in local spin density on the interfacial Ru atoms is evident, along with a 
  small local magnetic moment on the interfacial Ti atoms.}
\end{figure}

Since the magnetoelectric tensor, $\alpha$, is an intrinsic bulk quantity 
which relates bulk induced magnetization to applied electric field, it is not 
the appropriate parameter for describing the interfacial effects observed here.  
Instead, to characterize the magnetoelectric performance of the device, we start by introducing
the notion of \emph{spin-capacitance density} per unit area,  $\mathcal{C}_s$ which 
we define as
\begin{equation}
\mathcal{C}_s = \frac{\sigma_s}{V},
\end{equation}
where $\sigma_s$ is the amount of spin polarization per unit area induced 
by the voltage $V$.
This definition parallels that of the charge capacitance density per unit area,  
$\mathcal{C} = \frac{\sigma}{V}$, where $\sigma$ is the surface density of
free charge stored at the electrode.
Capacitance densities, however, depend both on the material properties and on the 
capacitor geometry, in particular, on the thickness of the dielectric film. 
Therefore, in order to obtain a parameter that characterizes the fundamental 
properties of the interface response, we further 
define the dimensionless parameter $\eta$, which is the ratio of spin capacitance 
to charge capacticance, i.e. $\eta = \mathcal{C}_s / \mathcal{C}$.
By definition, $\eta$ is also equal to the ratio of the induced spin with 
respect to the free charge accumulated at the capacitor plates, and
is clearly zero for non-magnetic electrodes; this interpretation provides a 
very intuitive picture of the interface response to a field.
The spin capacitance of our structure is $\mathcal{C}_s$=96.0 fF/$\mu$m$^{2}$, 
to be compared with a charge capacitance of $\mathcal{C}^0$=258 fF/$\mu$m$^2$, 
yielding a value of $\eta^0$=0.37. 

In order to investigate the origin of our calculated linear magnetoelectric effect,
we first compare this value of $\mathcal{C}^0$ (calculated here by using spin-polarized 
SRO electrodes) with the value obtained in Ref.~\onlinecite{Stengel/Spaldin_Nature:2006} 
for the same system calculated within the non-spin-polarized local density approximation.
Strikingly, the two capacitance values are identical within numerical accuracy, indicating
that the presence (or absence) of magnetic ordering in the electrode does not influence
the dielectric response of the capacitor. 
The static dielectric response of an insulator can be decomposed into an electronic and an
ionic contribution, where the latter is related to the frequency $\omega_i$
and the dipolar activity of the zone-center phonons:
\begin{equation}
\epsilon_0 = \epsilon_\infty + \frac{4\pi}{\Omega} \sum_i \frac{Z_i^2}{\omega_i^2} \quad .
\end{equation}
Here $\epsilon_\infty$ is the average frozen-ion permittivity of the supercell,
$\Omega$ is the volume and $Z_i$ are the ``mode effective charges'' obtained
by summing the ionic Born effective charges weighted by their relative contributions
to the normal mode eigenstates.
We found no appreciable differences in either $\epsilon_\infty$ or the
Born effective charges between the non-magnetic and the spin-polarized
calculations; the fact that the final value of $\mathcal{C}^0$ is
also unaffected strongly suggests that the \emph{frequencies} of the 
dipolar-active modes are themselves insensitive to magnetism.
Therefore, coupling between magnetism and phonons, which has been recognized as a 
viable route for magnetoelectric switching in perovskite materials~\cite{Fennie/Rabe:2006}, 
is not the driving force for the magnetoelectric response observed here, and a
different effect must be responsible.

To elucidate this behavior, we repeat our electric field calculations with the ions 
frozen in their initial centrosymmetric positions (Figure~\ref{hf_noavg}); 
this isolates the electronic response of the system and corresponds experimentally 
to the high-frequency limit.
Since by keeping the ions fixed we completely remove any structural effect on 
magnetism, any induced magnetization must be of intrinsic electronic origin, 
and can only be ascribed to the capacitive accumulation of spin-polarized carriers 
at the interface.
In this case we obtain an induced magnetic moment at each interface of 
$1.80 \times10^{-4}\ \mu_B$ per unit cell, corresponding to a spin-polarized charge 
density of 0.019~$\mu$C/cm$^{2}$, and a high frequency spin capacitance of 
$\mathcal{C}^{\infty}_s$= 6.99 fF/$\mu$m$^{2}$. All of these values are an order of
magnitude lower than the corresponding quantities in the static case.
Indeed the substantially lower induced magnetization and spin capacitance 
parallel an order of magnitude reduction in the charge capacitance, 
$\mathcal{C}^\infty$=20.3 fF/$\mu$m$^{2}$, which in turn derives from the 
suppressed dielectric response of the system when the ions are not allowed to 
relax. 
As a consequence, the corresponding value of $\eta^\infty$=0.34 is very close 
to that obtained in the static case.
The similarity in the values of $\eta$ calculated in the static and high-frequency 
limits indicates that the mechanisms leading to the screening of polar phonons 
in the static regime are the same as those screening electronic bound charges 
in the high-frequency regime, where changes in chemical bonding and/or structure
are not possible. Therefore we conclude that our system provides the first example
of a {\it carrier-mediated magnetoelectric effect}, which results entirely from the
capacitive accumulation of spin-polarized carriers at the interfaces. 

\begin{figure}
  \centering
  \includegraphics[width=0.950\textwidth]{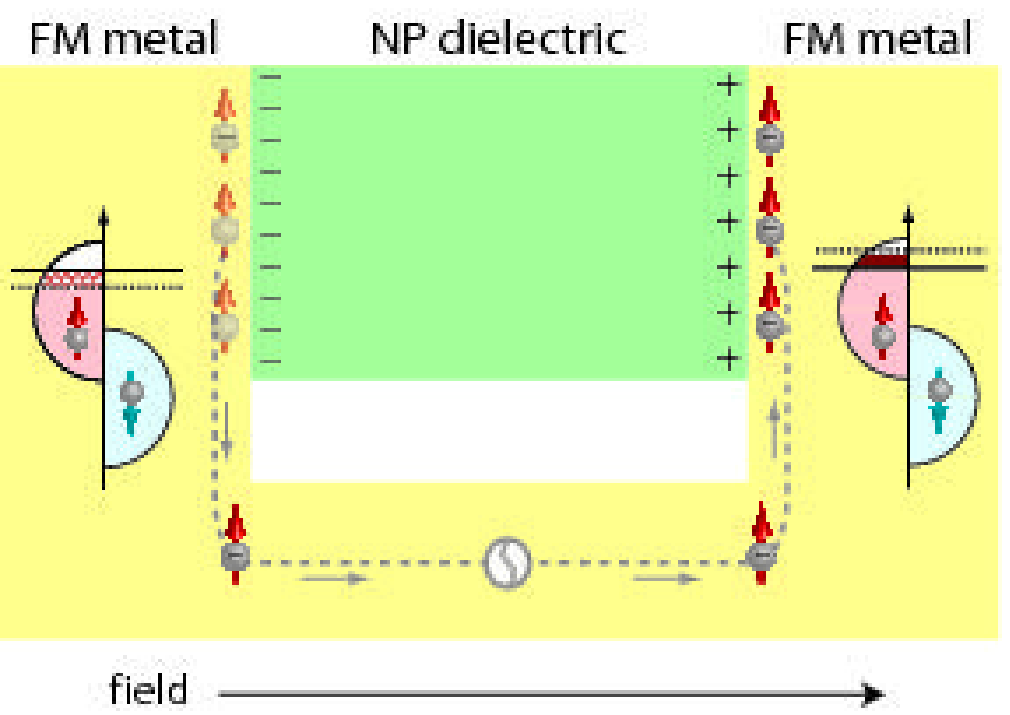}
  \caption{\label{CM-cartoon}(color online)~Schematic of the mechanism causing the induced
magnetization at a ferromagnetic (FM) metal -- non-polar (NP) dielectric interface. The 
inset density of states cartoons show the changes to the occupation of half-metallic 
densities of 
states in the electrodes when the current flows in response to the applied electric 
field. The accumulation of up-spin electrons adjacent to the positive face of the 
dielectric, and their depletion from the negative face, leads to the net magnetization
shown in Figs.\ {\protect \ref{spin_capacitor_fig} and \ref{hf_noavg}}.}
\end{figure}
The origin of the carrier-mediated magnetoelectric effect is shown schematically
in Figure~\ref{CM-cartoon}. 
On application of an external field, free carriers accumulate at the 
capacitor plates, which are partially screened by the dielectric polarization
of the STO film.
In the half-metallic limit all displaced electrons are spin-polarized 
in the same direction (up in the figure); in the present case there is 
a partial cancellation between spin-up and spin-down carriers that 
reflects the incomplete spin polarization at the Fermi level of the 
interfacial SRO layer. 
This process accumulates up-spin magnetization adjacent to the positively 
charged electrode, leaving behind an absence of up-spin magnetization 
(or equivalently down-spin polarized holes) at the negative plate. 

One possible mechanism for detecting the induced magnetization is the 
magneto-optical Kerr effect.
Indeed, we note that the recent report of enhanced Kerr response in a 2~nm 
thick FePt film when an electric field is applied across an electrochemical  
cell with FePt electrodes \cite{Weisheit_et_al:2007} is also likely a 
manifestation of the carrier-mediated magnetoelectric effect that we describe 
here.
In Ref.~\onlinecite{Weisheit_et_al:2007} a Kerr rotation of $\sim$1$^{\circ}$ 
was obtained for an applied potential of 600 mV, which generated 0.015 
electrons per unit cell; in our case an equivalent potential would induce a 
spin polarization of  0.05 spin-polarized electrons per unit cell, which
should therefore be readily detectable. 

Substitution of our field-polarized dielectric by a ferroelectric with a spontaneous 
electrical polarization would yield a correspondingly larger spin density and 
interfacial magnetoelectric effect as well as an added degree of freedom from the 
switchable polarization. 
The parameter $\eta$ has a particularly transparent meaning in this case, relating
the accumulated spin at the interface to the bulk polarization of the ferroelectric,
$P$:
Neglecting the effect of depolarizing fields (which is a good approximation in
the limit of a sufficiently thick ferroelectric film), the free charge accumulated 
at the electrode exactly compensates the surface polarization charge induced by the 
spontaneous electric displacement. 
Therefore, $\eta=\sigma_s/P$, and the induced spin at the interface
can be readily predicted by the knowledge of $\eta$ and $P$, under the
assumption that the behavior of the free carriers is linear in $P$.
For example, from the numbers calculated here we predict $\sigma_s$=12 
$\mu$C/cm$^{2}$ for a SrRuO$_3$/BaTiO$_3$ interface 
($P_{BTO}$=32~$\mu$C/cm$^{2}$); this corresponds to a fully switchable 
magnetic moment of 0.11 $\mu_B$.
To substantiate these arguments we repeated the calculation by replacing STO
with BTO~\footnote{Six Sr atoms within the insulating layer were replaced with
  Ba, which corresponds to assuming a TiO$_2$ (SrO) termination of the BTO (SRO)
  lattice. The in-plane lattice parameter was kept fixed to the calculated bulk
  STO value, while both out-of-plane lattice parameter and atomic positions were
  fully relaxed while imposing mirror symmetry with respect to a plane parallel
  to the interface.}.
We restricted our analysis to the high-frequency case to avoid complications
related to the proper treatment of BTO lattice instabilities; by analogy with 
the SRO/STO case, it is unlikely that the fully relaxed static behavior will 
significantly differ.
The resulting $\eta^{BTO}_\infty=0.37$ is indeed very close to the SRO/STO value,
confirming the validity of our arguments.
In addition, such ferroelectric/ferromagnetic metal heterostructures would exhibit a new
type of {\it interfacial multiferroism}, resulting from the simultaneous small magnetization 
of the ferroelectric by the magnetic metal reported in this work, and the polarization 
of the metallic magnet by the depolarizing field of 
the ferroelectric described previously in Ref.~\onlinecite{Stengel/Spaldin_Nature:2006}. 
We note that the interfacial multiferroism results from the attempt by 
the metallic electrode to screen the capacitive charge at the interface, and 
is mediated by the capacitive accumlation of spin-polarized carriers at the 
interfacial layers. The behavior is therefore distinct from that in
composite multiferroics, which combine ferro- or ferri-magnetic {\it insulators} 
with ferroelectrics (see for example Refs. 
\onlinecite{Srinivasan_et_al:2001,Dong_et_al:2003,Zheng_et_al:2004}).

This class of heterostructures was recently investigated by Duan \textit{et al.} 
\cite{Duan/Jaswal/Tsymbal:2006}, who calculated changes in magnetism induced by
ferroelectric displacements at the interface between ferroelectric BaTiO$_3$ and 
ferromagnetic Fe.
According to Ref.~\onlinecite{Duan/Jaswal/Tsymbal:2006}, the positively and
negatively polarized faces of the BaTiO$_3$ film have magnetic moments at the 
interface with the Fe electrodes which differ by almost 0.3 $\mu_B$ per surface
unit cell; the differences are ascribed to changes in interfacial chemical bonding 
and hybridization. While the magnitude of this value is consistent with our estimates, 
surprisingly
most of the change in moment in Ref.~\onlinecite{Duan/Jaswal/Tsymbal:2006} resulted
from changes in spin polarization of the Ti $d$ states, which we find to be minimally affected
by the electric field.
We note, however, that such changes in hybridization are strongly dependent on the position 
of Ti $d$ states with respect to the Fermi level and can be affected by the well-known
underestimation of the fundamental insulating gap that plagues most commonly
used DFT functionals.
Conversely, the carrier-mediated accumulation of spin-polarized carriers investigated 
here is a general characteristic of all ferromagnetic electrodes in contact with insulators, 
regardless of the details of the bonding; since its magnitude is similar to hybridization 
effects, it should not be neglected.

Finally, we mention some possible related applications that could exploit
the predicted carrier-mediated magnetoelectric behavior.
With the ability to capacitively store spin-polarized charge, we see that this 
heterostructure provides the spintronic analogue of a thin film charge 
capacitor, and could therefore find application in spin-polarized extensions to regular 
capacitor applications, such as a filter of spin-polarized direct current, or 
in spin logic circuits. 
In addition, since the spin capacitance mechanism only allows current to flow when the 
leads are magnetized in the same direction, the capacitor could be used as a sensitive 
magnetoresistive detector with the desirable feature that the electrons never cross 
interfaces and so are less susceptible to scattering of their spin orientation.  
Since the change in {\it total} magnetization is small this material combination is
not immediately transferable to electric field tunable inductors. The magnitude of 
the effect would be enhanced, however, by the use of materials with higher spin 
polarization at the Fermi surface, and lower overall magnetization; in particular 
half-metallic antiferromagnetic electrodes \cite{Pickett:1998} with zero net 
magnetization would give an infinite relative change in net magnetization with 
electric field. 
We note also that the calculated LSDA dielectric constant of SrTiO$_3$ is 490, 
whereas experimental values can be tens of thousands, and the induced magnetization 
will scale accordingly. 

To summarize, we have used first-principles density functional calculations 
combined with finite electric field methods to demonstrate a linear 
magnetoelectric response in a complex oxide heterostructure. 
This new carrier-mediated mechanism adds an additional degree of freedom in the 
design and functionalization of magnetoelectric multiferroic materials, and may 
guide the way forward in designing electric-field switchable magnetic devices.
With growing interest in superlattice multiferroic composites, the phenomenon we 
describe may elucidate experimentally observed effects that to date are 
attributed to unknown interfacial responses to applied electric 
(and magnetic) fields \cite{Chaudhuri_et_al:2007}.
Our results also expand the growing body of literature demonstrating the 
novel functionality that can be achieved at oxide interfaces 
\cite{Ohtomo_et_al:2002,
Ohtomo/Hwang:2004,
Thiel_et_al:2006,
Hwang_science:2006,
Huijben_et_al:2006,
Yamada_et_al:2004,
de_Teresa_et_al:1999,
Lottermoser_et_al:2004} 
and suggest an additional route to novel oxide-based interfacial devices
 \cite{Kroemer_nobel:2001}.

This work was supported by the DOE SciDac program on Quantum Simulations of 
Materials and Nanostructures, grant number DE-FC02-06ER25794 (M.S.), and by 
the NSF NIRT program, grant number 0609377 (J.M.R). 
N.S. thanks the Miller Institute at UC Berkeley for their
support through a Miller Research Professorship.

\end{document}